\documentclass[final,3p]{elsarticle}

\usepackage{amssymb}
\usepackage{amsmath}
\usepackage{graphicx}
\usepackage[usenames]{color}

\journal{Journal of Subatomic Particles and Cosmology}
\graphicspath{{./Figs_QCS2026_proceedings/}}

\begin{document}

\begin{frontmatter}

\title{In-Medium Modification of $\phi$ Meson Mass over Temperature and Momentum}

\author[zju]{Hidefumi Matsuda}
\ead{da.matsu.00.bbb.kobe@gmail.com}
\author[jaea]{Philipp Gubler}
\ead{gubler.philipp@jaea.go.jp; philipp.gubler1@gmail.com}
\author[zju,rcnp]{Koichi Hattori}
\ead{koichi.hattori@zju.edu.cn}

\affiliation[zju]{organization={Zhejiang Institute of Modern Physics, Department of Physics, Zhejiang University},
           city={Hangzhou},
           postcode={310027},
           country={China}}
\affiliation[jaea]{organization={Advanced Science Research Center, Japan Atomic Energy Agency},
           city={Tokai},
           postcode={319-1195},
           country={Japan}}
\affiliation[rcnp]{organization={Research Center for Nuclear Physics (RCNP), Osaka University},
           city={Osaka},
           postcode={567-0047},
           country={Japan}}

\begin{abstract}
We analyze the in-medium modification of the $\phi$ meson mass below 
the pseudo-critical temperature $T_\text{c}$ 
at vanishing baryon chemical potential using QCD sum rules. The sum rules are applied separately to the transverse and longitudinal polarization modes of the $\phi$ meson defined relative to the spatial momentum in the medium rest frame. We map out the temperature and momentum dependence of the mass in each mode and quantify the resulting transverse--longitudinal mass splitting. The splitting develops with increasing temperature and momentum.
\end{abstract}


\end{frontmatter}

\section{Introduction}
\label{sec:introduction}

Vector mesons such as the $\rho$ and $\phi$ are finite-lifetime excitations propagating through hadronic matter. Their properties are modified through interactions with the surrounding medium.
Changes in their spectral properties, including their masses, decay widths, and decay constants, reflect properties of the medium, such as the partial restoration of chiral symmetry and thermal fluctuations. For this reason, in-medium modifications of vector mesons have long been studied, both theoretically and experimentally, as probes of hadronic matter at finite temperature and density. Among them, the $\phi$ meson has attracted particular attention because its coupling to the medium through the strange-quark channel allows it to serve as a distinctive probe. 

Of particular interest in the present study is the splitting between the longitudinal and transverse polarization modes at finite momentum. In vacuum, Lorentz symmetry constrains the vector meson spectrum to depend only on the invariant momentum $q^2$. A thermal medium, by contrast, selects a preferred rest frame $u^\mu$, thereby partially breaking Lorentz symmetry. At nonzero spatial momentum in this frame, the vector meson correlator is decomposed into two independent polarization components, longitudinal and transverse with respect to the spatial momentum. 
The spectrum of each component depends not only on $q^2$ but also on the squared spatial momentum $v^2=(u\cdot q)^2-q^2$, and the longitudinal and transverse masses can in general vary differently with temperature and momentum. This splitting is a direct manifestation of the Lorentz symmetry breaking induced by the medium.

In this work, we investigate the longitudinal and transverse $\phi$ meson masses at finite temperature and spatial momentum in a low-temperature hadronic medium at vanishing baryon chemical potential, using QCD sum rules.
This is the first attempt to investigate the $\phi$-meson splitting at finite temperature and finite momentum within the framework of QCD sum rules, while such a splitting has been studied at finite density in Refs.~\cite{Lee:1997zta,Kim:2019ybi}. We consider temperatures up to $T=150~\mathrm{MeV}$, which is close to the pseudo-critical temperature $T_{\mathrm c}$, and momenta up to $2~\mathrm{GeV}$. For the analysis based on QCD sum rules, we construct the operator product expansion (OPE) following Ref.~\cite{Kim:2017nyg}, including condensate contributions up to dimension six. The non-scalar condensates in the hadronic medium at finite temperature are evaluated within the dilute pion-gas approximation~\cite{Hatsuda:1992bv}. Further details on the theoretical formulation and numerical implementation can be found in our full paper~\cite{Matsuda:2026kaz}.

\section{QCD Sum Rules for the $\phi$ Meson at Finite Temperature}
\label{sec:qcd_sum_rule_framework}
QCD sum rules provide a framework for extracting hadronic spectral properties by matching two representations of the same current correlation function. In the deep Euclidean region, the correlator is directly evaluated using the OPE, while the dispersion relation expresses the same correlator as an integral over the spectral function, which is modeled phenomenologically. Matching the two representations then constrains the spectral properties.

We study the $\phi$ meson through the thermal correlation function of the strange vector current,
\begin{align}
\Pi^{\mu\nu}(q,T)
&=
i\int d^4x\,e^{iq\cdot x}
\left\langle
\mathcal{T}J^\mu(x)J^\nu(0)
\right\rangle_T\ ,
\qquad
J^\mu=\bar{s}\gamma^\mu s\ ,
\label{eq:thermal_correlator_proceedings}
\end{align}
where $\langle\cdots\rangle_T$ denotes the thermal expectation value. The correlator is symmetric under the exchange of $\mu$ and $\nu$ in 
the absence of parity violation or external fields, and satisfies the 
Ward identity, $q_\mu\Pi^{\mu\nu}=0$. In vacuum, Lorentz symmetry reduces the correlator to a single scalar amplitude,
$\Pi^{\mu\nu} = P^{\mu\nu} \Pi^{\mathrm{vac}}$, with $P^{\mu\nu} = q^\mu q^\nu - q^2 g^{\mu\nu}$.
At finite temperature, the medium four-velocity $u^\mu$ provides an additional four-vector, 
inducing two independent tensor structures,
\begin{align}
\Pi^{\mu\nu}(q,T)
=
P_{\mathrm T}^{\mu\nu}
\Pi_{\mathrm T}(q^2,v^2;T)
+
P_{\mathrm L}^{\mu\nu}
\Pi_{\mathrm L}(q^2,v^2;T)\ .
\label{eq:TL_decomposition_proceedings}
\end{align}
Here $P_{\mathrm T}^{\mu\nu}$ is orthogonal to the spatial momentum, satisfying $\{q_\mu-u_\mu (u\cdot q)\}P_{\mathrm T}^{\mu\nu}=0$, whereas $P_{\mathrm L}^{\mu\nu}$ is not. These two modes are thus identified as the transverse and longitudinal modes with respect to the spatial momentum.
In the limit where the momentum is parallel to the medium four-velocity, $q^\mu \propto u^\mu$, the two scalar amplitudes become degenerate, i.e., $\Pi_{\mathrm T}=\Pi_{\mathrm L}$.
For each polarization channel $X=\mathrm L,\mathrm T$, the scalar correlation function is related to the corresponding spectral function through the dispersion relation
\begin{align}
\Pi_X(q^2,v^2;T)
=
\int_{-v^2}^{\infty}
ds\,
\frac{\rho_X(s,v^2;T)}{s-q^2}\ ,
\qquad
(X=\mathrm L,\mathrm T)\ ,
\label{eq:dispersion_relation_proceedings}
\end{align}
where we omitted the
subtraction terms that vanish under the Borel transformation introduced below. 

In the deep Euclidean region, $Q^2\equiv -q^2\gg\Lambda_{\mathrm{QCD}}^2$, the left-hand side of Eq.~\eqref{eq:dispersion_relation_proceedings} is evaluated using the OPE expression $\Pi_X^{\mathrm{OPE}}(q^2,v^2;T)$.
We include scalar and non-scalar condensates up to dimension six, together with the Wilson coefficients derived in Ref.~\cite{Kim:2017nyg}. 
The complete OPE expressions, numerical input parameters, and condensates used in the analysis are shown in Ref.~\cite{Matsuda:2026kaz}. It should be noted that the non-scalar condensates vanish in vacuum because of the Lorentz symmetry and are induced entirely by the medium. Their thermal expectation values are evaluated within the dilute pion-gas approximation through the one-pion matrix elements, as discussed in Refs.~\cite{Hatsuda:1992bv,Matsuda:2026kaz,Gubler:2018ctz}.

The spectral function in each polarization channel, appearing on the right-hand side of Eq.~\eqref{eq:dispersion_relation_proceedings}, is modeled as the sum of a $\phi$-meson pole and a contribution from higher-energy states,
\begin{align}
\rho_X(s,v^2;T)
=
f_{\phi,X}^2(T,v^2)
\delta\!\left(
s-m_{\phi,X}^2(T,v^2)
\right)
+
\rho_{\mathrm{pert}}(s)\,
\theta\!\left(
s-s_{0,X}(T,v^2)
\right)\ ,
\label{eq:spectral_ansatz_proceedings}
\end{align}
where $m_{\phi,X}$ and $f_{\phi,X}$ denote the polarization-dependent mass and decay constant, respectively, and $s_{0,X}$ is the threshold marking the onset of continuous higher-energy states.

We apply the Borel transformation to both sides of the dispersion relation 
to effectively carry out the matching between the OPE representation of the correlation function and the phenomenologically parametrized spectral function.
The Borel transformation is defined by
\begin{align}
  \mathcal{B}_{M^2}\left[f\right]
  &\equiv
  \lim_{\substack{Q^2,N\to\infty\\ Q^2/N=M^2}}
  \frac{(Q^2)^N}{(N-1)!}
  \left(-\frac{d}{dQ^2}\right)^N f(Q^2)\ ,
\label{eq:borel_operator_definition}
\end{align}
and transforms the factor $1/(s-q^2)$ into $(1/M^2)e^{-s/M^2}$. Thus, the dispersion relation becomes
\begin{align}
\mathcal{B}_{M^2}
\!\left[
\Pi_X^{\mathrm{OPE}}(q^2,v^2;T)
\right]
=
\frac{1}{M^2}
\int_{-v^2}^{\infty}
ds\,
\rho_X(s,v^2;T)
e^{-s/M^2}\ ,\qquad
X=\mathrm{L},\mathrm{T}\ .
\label{eq:borel_dispersion_relation_proceedings}
\end{align}
Substituting the ansatz of Eq.~\eqref{eq:spectral_ansatz_proceedings} into Eq.~\eqref{eq:borel_dispersion_relation_proceedings} gives
\begin{align}
M^2
\mathcal{B}_{M^2}
\!\left[
\Pi_X^{\mathrm{OPE}} (q^2,v^2;T)
\right]
=
f_{\phi,X}^2(T,v^2)
\exp\!\left[
-\frac{m_{\phi,X}^2(T,v^2)}{M^2}
\right]
+
\int_{s_{0,X}}^{\infty}
ds\,
\rho_{\mathrm{pert}}(s)
e^{-s/M^2}\ .
\label{eq:borel_pole_continuum_proceedings}
\end{align}
The exponential factor suppresses the contribution of high-energy states and enhances the relative sensitivity to the low-lying $\phi$-meson resonance.
We define
\begin{align}
\mathcal{R}_X(M^2,v^2;T,s_{0,X})
&\equiv
M^2
\mathcal{B}_{M^2}
\!\left[
\Pi_X^{\mathrm{OPE}} (q^2,v^2;T)
\right]
-
\int_{s_{0,X}}^{\infty}
ds\,
\rho_{\mathrm{pert}}(s)
e^{-s/M^2}\ ,
\label{eq:borel_moment_R_proceedings}
\end{align}
which satisfies
\begin{align}
\mathcal{R}_X(M^2,v^2;T,s_{0,X})
=
f_{\phi,X}^2(T,v^2)
\exp\!\left[
-\frac{m_{\phi,X}^2(T,v^2)}{M^2}
\right]\ .
\label{eq:continuum_subtracted_sum_rule_proceedings}
\end{align}
Taking the logarithm of Eq.~\eqref{eq:continuum_subtracted_sum_rule_proceedings} and then differentiating it with respect to $M^{-2}$ gives the polarization-dependent mass,
\begin{align}
m_{\phi,X}^2(M^2,v^2;T,s_{0,X})
=
-
\frac{d}{d(M^{-2})}
\ln
\mathcal{R}_X(M^2,v^2;T,s_{0,X}).
\label{eq:mass_extraction_proceedings}
\end{align}
Once the mass is determined, the decay constant follows from
\begin{align}
f_{\phi,X}^2(M^2,v^2;T,s_{0,X})
=
\mathcal{R}_X(M^2,v^2;T,s_{0,X})
\exp\!\left[
\frac{m_{\phi,X}^2(M^2,v^2;T,s_{0,X})}{M^2}
\right].
\label{eq:decay_constant_extraction_proceedings}
\end{align}
The extracted quantities $m^2_{\phi,X}$ and $f^2_{\phi,X}$ retain a residual dependence on the Borel mass $M$ and the threshold $s_{0,X}$.
The analysis is therefore restricted to a Borel window,
\begin{align}
M_{\min,X} < M_X < M_{\max,X}\ .
\end{align}
The lower bound $M_{\min,X}$ is set by a condition on the convergence of the OPE, while the upper bound $M_{\max,X}$ is set by a condition that the pole contribution remains sufficiently large relative to the contribution from higher-energy states.
The upper bound $M_{\max,X}$ depends on the continuum threshold $s_{0,X}$ and is therefore written as $M_{\max,X}(s_{0,X})$.
For each temperature, momentum, and polarization channel, the continuum threshold is varied over the range in which a valid Borel window exists.
The optimal value of $s_{0,X}$ is chosen so that the extracted mass $m_{\phi,X}$ is least sensitive to $M$ over the corresponding Borel window, $M_{\min,X}<M_X<M_{\max,X}(s_{0,X})$.
The final mass and decay constant are obtained by averaging over the Borel window corresponding to the optimized $s^0$. The explicit Borel-transformed OPE and the quantitative criteria defining the Borel window are given in Ref.~\cite{Matsuda:2026kaz}.
 
\section{Mass map in the \((T,v)\) plane}
\label{sec:mass_map}

In this section, we present the temperature and spatial-momentum dependence of the $\phi$-meson masses over the ranges $0\leq T\leq150~\mathrm{MeV}$ and $0\leq v\leq2~\mathrm{GeV}$. We first compare the transverse and longitudinal masses, $m_{\phi,\mathrm T}(T,v)$ and $m_{\phi,\mathrm L}(T,v)$, and then examine their difference, $\Delta m_\phi(T,v)\equiv m_{\phi,\mathrm T}(T,v)-m_{\phi,\mathrm L}(T,v)$. 

\begin{figure}[!htbp]
\centering
\begin{minipage}{0.46\textwidth}
\centering
\includegraphics[width=\linewidth]{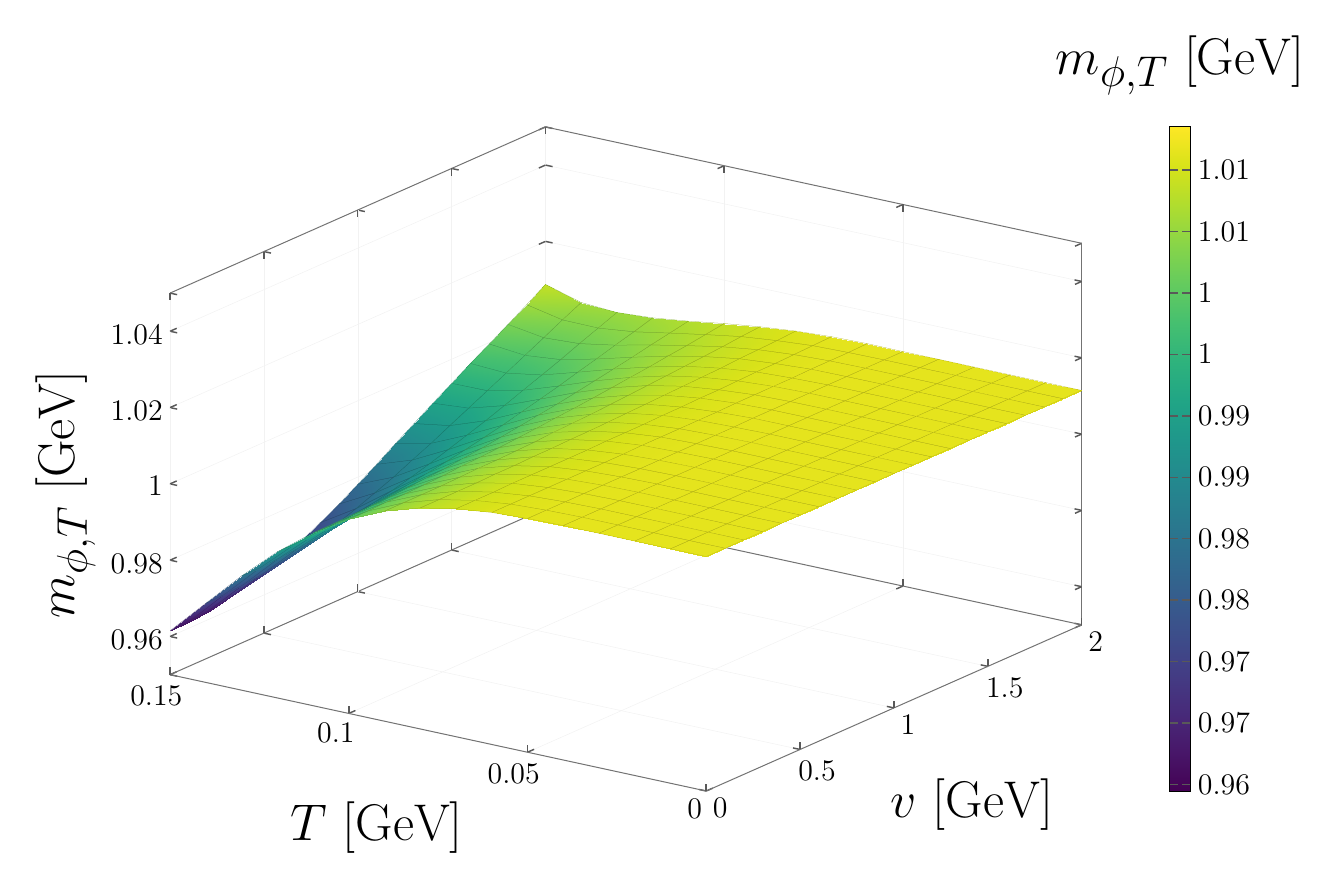}
\par\vspace{1mm}
{\small (a) Transverse channel}
\end{minipage}
\hspace{0.03\textwidth}
\begin{minipage}{0.46\textwidth}
\centering
\includegraphics[width=\linewidth]{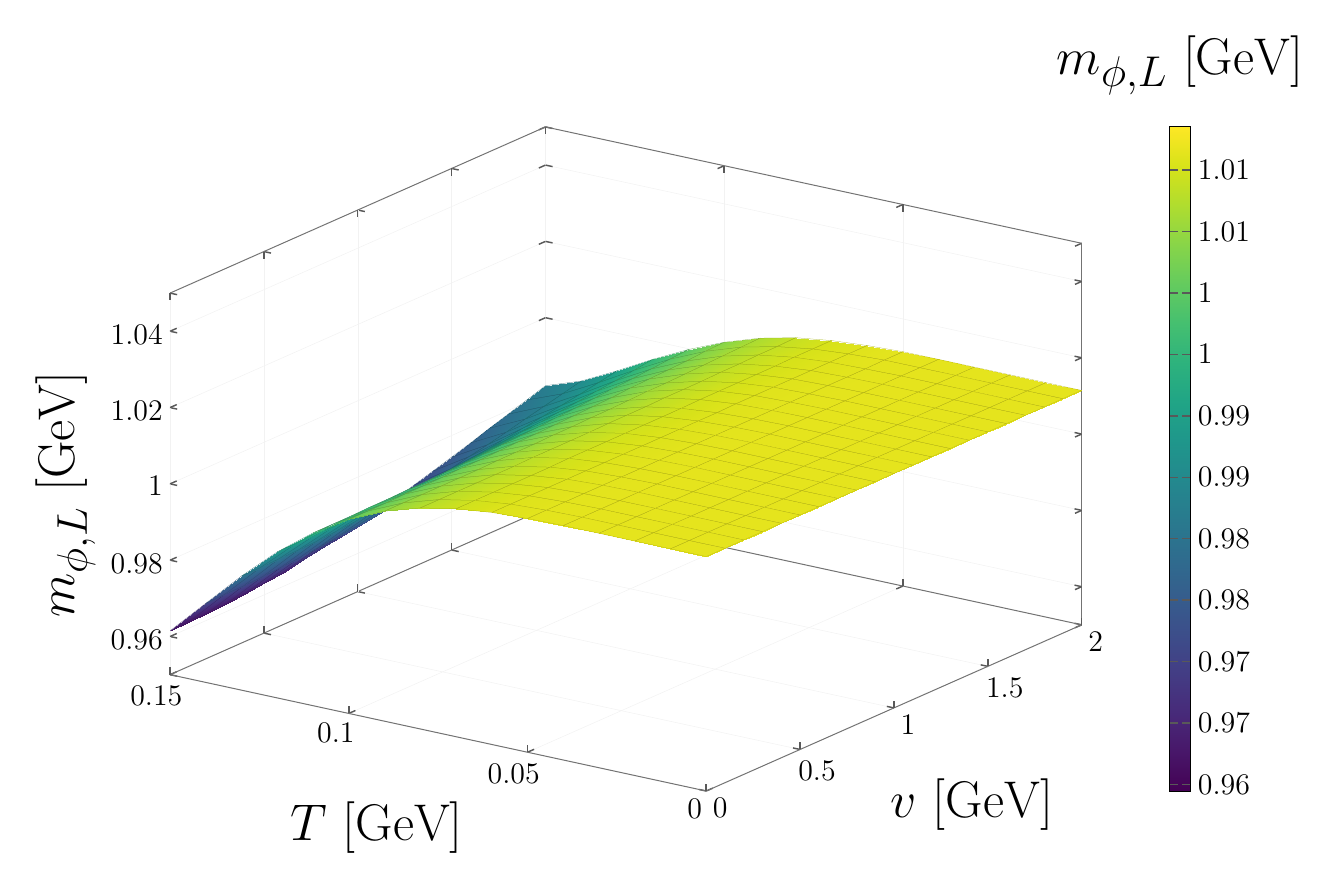}
\par\vspace{1mm}
{\small (b) Longitudinal channel}
\end{minipage}
\caption{
Three-dimensional representation of the $\phi$-meson mass over the temperature and momentum plane.
}
\label{fig:mphi_TL_maps}
\end{figure}

Figure~\ref{fig:mphi_TL_maps} shows 3D plots of the transverse and longitudinal masses
as functions of temperature and spatial momentum.
At zero spatial momentum, the two masses coincide, as required by the rotational symmetry in the medium rest frame. In this limit, the mass stays nearly constant at low temperature and begins to decrease above roughly $T=60~\mathrm{MeV}$. At fixed temperature, both masses increase with momentum, but the transverse mass rises faster than the longitudinal one. As a result, the two surfaces gradually separate as the momentum grows, with $m_{\phi,\mathrm T}>m_{\phi,\mathrm L}$.

\begin{figure}[!htbp]
  \centering
  \includegraphics[width=0.52\textwidth]{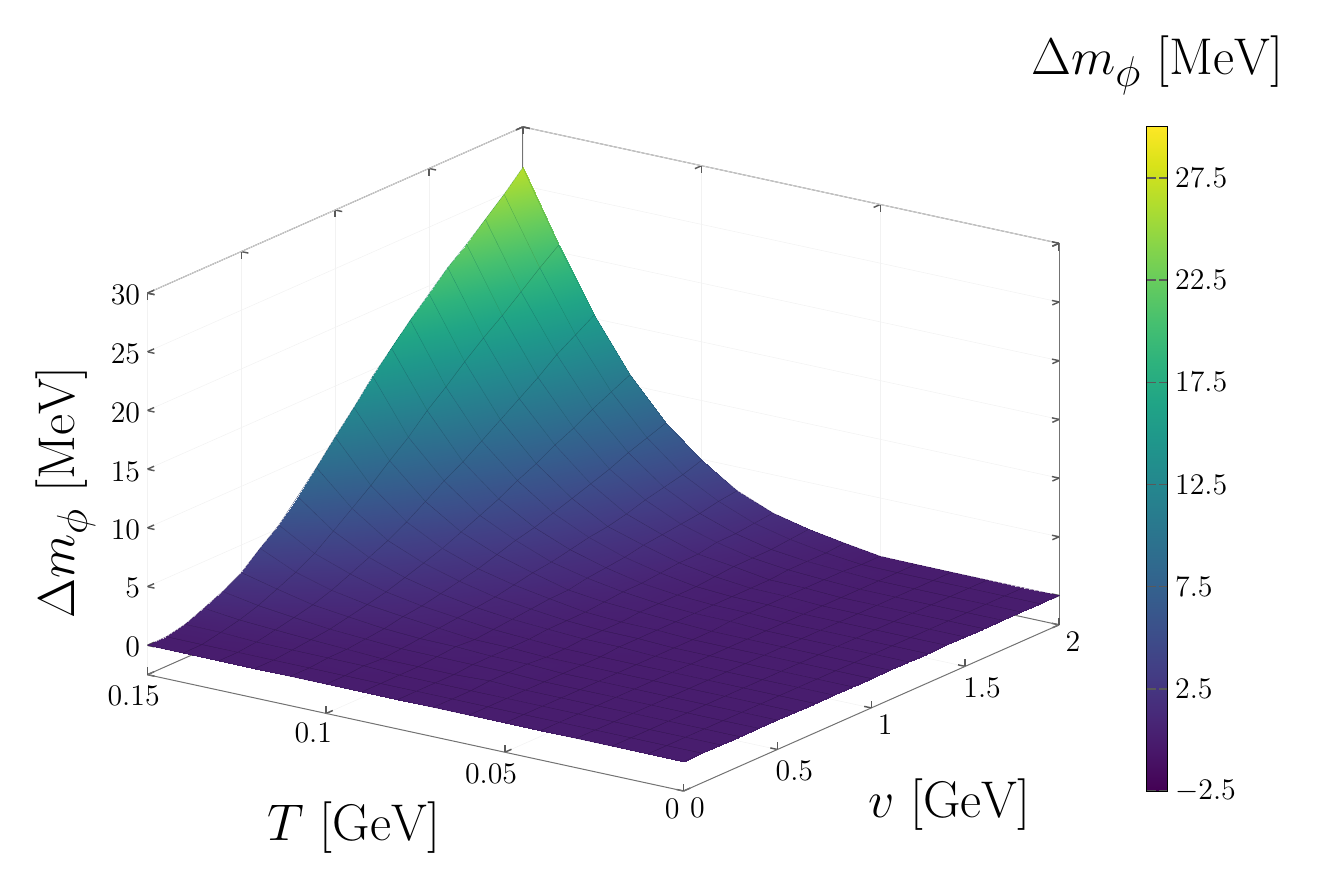}
  \caption{Three-dimensional representation of the transverse-longitudinal mass splitting \(\Delta m_\phi(T,v)\equiv m_{\phi,T}(T,v)-m_{\phi,L}(T,v)\). This plot visualizes how the polarization dependence develops over the temperature and momentum plane.}
  \label{fig:delta_mphi_map}
\end{figure}

Figure~\ref{fig:delta_mphi_map} isolates the polarization dependence of the $\phi$-meson mass with a 3D plot of $\Delta m_\phi(T,v)$, the difference between the transverse and longitudinal masses.
The splitting vanishes at zero spatial momentum and grows with both temperature and spatial momentum, reaching about $26~\mathrm{MeV}$ near the upper limits of the temperature and momentum ranges considered here.
As discussed in detail in Ref.~\cite{Matsuda:2026kaz}, this splitting originates predominantly from the medium-induced non-scalar condensates in the OPE, with the leading dimension-four condensates $F$ and $G_2$ giving the dominant contributions.

\section{Summary}
\label{sec:summary}

We have investigated the temperature and spatial-momentum dependence of the transverse and longitudinal $\phi$-meson masses in the low-temperature hadronic medium at vanishing baryon density using QCD sum rules. At zero spatial momentum, the two modes are degenerate and their common mass decreases above approximately $T=60~\mathrm{MeV}$. At finite momentum, both masses increase, with the transverse mass exceeding the longitudinal one. The resulting mass splitting grows with temperature and momentum and reaches about $26~\mathrm{MeV}$ within the range considered. 
We conclude that the polarization dependence, allowed at finite temperature and momentum, indeed manifests itself in the nonzero mass splitting.

\section*{Acknowledgments}

This work is supported by the National Natural Science Foundation of China (NSFC) under grant numbers W2433010 and W2532002 and 
by JSPS KAKENHI under grant numbers JP25H00400, JP26K07113, and 23K22487.

\bibliographystyle{elsarticle-num}
\bibliography{apssamp}

@article{Lee:1997zta,
    author = "Lee, Su Houng",
    title = "{Vector mesons in-medium with finite three momentum}",
    eprint = "nucl-th/9705048",
    archivePrefix = "arXiv",
    reportNumber = "SNUTP-97-068",
    doi = "10.1103/PhysRevC.57.927",
    journal = "Phys. Rev. C",
    volume = "57",
    pages = "927--930",
    year = "1998",
    note = "[Erratum: Phys.Rev.C 58, 3771 (1998)]"
}

@article{Kim:2019ybi,
    author = "Kim, HyungJoo and Gubler, Philipp",
    title = "{The {\ensuremath{\phi}} meson with finite momentum in a dense medium}",
    eprint = "1911.08737",
    archivePrefix = "arXiv",
    primaryClass = "hep-ph",
    doi = "10.1016/j.physletb.2020.135412",
    journal = "Phys. Lett. B",
    volume = "805",
    pages = "135412",
    year = "2020"
}

@article{Kim:2017nyg,
    author = "Kim, HyungJoo and Gubler, Philipp and Lee, Su Houng",
    title = "{Light vector correlator in medium: Wilson coefficients up to dimension 6 operators}",
    eprint = "1703.04848",
    archivePrefix = "arXiv",
    primaryClass = "hep-ph",
    doi = "10.1016/j.physletb.2017.06.047",
    journal = "Phys. Lett. B",
    volume = "772",
    pages = "194--199",
    year = "2017",
    note = "[Erratum: Phys.Lett.B 779, 498--498 (2018)]"
}

@article{Hatsuda:1992bv,
    author = "Hatsuda, Tetsuo and Koike, Yuji and Lee, Su-Houng",
    title = "{Finite temperature QCD sum rules reexamined: rho, omega and A1 mesons}",
    reportNumber = "UMD-PP-92-203, DOE-ER-40322-156, YSTP-92-011",
    doi = "10.1016/0550-3213(93)90107-Z",
    journal = "Nucl. Phys. B",
    volume = "394",
    pages = "221--266",
    year = "1993"
}

@article{Gubler:2018ctz,
    author = "Gubler, Philipp and Satow, Daisuke",
    title = "{Recent Progress in QCD Condensate Evaluations and Sum Rules}",
    eprint = "1812.00385",
    archivePrefix = "arXiv",
    primaryClass = "hep-ph",
    doi = "10.1016/j.ppnp.2019.02.005",
    journal = "Prog. Part. Nucl. Phys.",
    volume = "106",
    pages = "1--67",
    year = "2019"
}

@article{Matsuda:2026kaz,
    author = "Matsuda, Hidefumi and Gubler, Philipp and Hattori, Koichi",
    title = "{Polarization dependence of the $\phi$ meson from finite-temperature QCD sum rules}",
    eprint = "2605.27347",
    archivePrefix = "arXiv",
    primaryClass = "hep-ph",
    month = "5",
    year = "2026"
}

\end{document}